\documentclass[pdflatex,sn-mathphys-num]{sn-jnl}

\usepackage{graphicx}%
\usepackage{multirow}%
\usepackage{amsmath,amssymb,amsfonts}%
\usepackage{amsthm}%
\usepackage{mathrsfs}%
\usepackage[title]{appendix}%
\usepackage{xcolor}%
\usepackage{textcomp}%
\usepackage{manyfoot}%
\usepackage{booktabs}%
\usepackage{algorithm}%
\usepackage{algorithmicx}%
\usepackage{algpseudocode}%
\usepackage{listings}%

\theoremstyle{thmstyleone}%
\theoremstyle{thmstyletwo}%

\theoremstyle{thmstylethree}%

\begin{document}

\title[Article Title]{40 Years of Interdisciplinary Research: Phases, Origins, and Key Turning Points (1981–2020)}


\author[1,2]{\fnm{Guoyang} \sur{Rong}}\email{chrisr@whu.edu.cn}

\author*[2,3]{\fnm{Ying} \sur{Chen}}\email{matcheny@nus.edu.sg}

\author*[1]{\fnm{Feicheng} \sur{Ma}}\email{fchma@whu.edu.cn}

\author[4,5]{\fnm{Thorsten} \sur{Koch}}\email{koch@zib.de}

\affil[1]{\orgdiv{School of Information Management}, \orgname{Wuhan University}, \orgaddress{\city{Wuhan}, \postcode{430072}, \country{China}}}

\affil[2]{\orgdiv{Center for Quantitative Finance, Department of Mathematics}, \orgname{National University of Singapore}, \orgaddress{\postcode{119076}, \country{Singapore}}}

\affil[3]{\orgdiv{Risk Management Institute}, \orgname{National University of Singapore}, \orgaddress{\postcode{119076}, \country{Singapore}}}

\affil[4]{\orgdiv{Institute of Mathematics}, \orgname{Technische Universität Berlin}, \orgaddress{\city{Berlin}, \postcode{10623}, \country{Germany}}}

\affil[5]{\orgdiv{Applied Algorithmic Intelligence Methods}, \orgname{Zuse Institute Berlin}, \orgaddress{\city{Berlin}, \postcode{14195}, \country{Germany}}}


\abstract{This study examines the historical evolution of interdisciplinary research (IDR) over a 40-year period, focusing on its dynamic trends, phases, and key turning points. 
We apply time series analysis to identify \textit{critical years for interdisciplinary citations} (CYICs) and categorizes IDR into three distinct phases based on these trends: Period I (1981–2002), marked by sporadic and limited interdisciplinary activity; Period II (2003–2016), characterized by the emergence of large-scale IDR led primarily by Medicine, with significant breakthroughs in cloning and medical technology; and Period III (2017–), where IDR became a widely adopted research paradigm. 
Our findings indicate that IDR has been predominantly concentrated within the Natural Sciences, with Medicine consistently at the forefront, and highlights increasing contributions from Engineering and Environmental disciplines as a new trend. 
These insights enhance the understanding of the evolution of IDR, its driving factors, and the shifts in the focus of interdisciplinary.}




\maketitle

\section{Introduction}\label{sec1}

Interdisciplinary Research (IDR) is a mode of research that integrates information, data, techniques, tools, perspectives, concepts, and/or theories from two or more disciplines or bodies of specialised knowledge to advance fundamental understanding or to solve problems \cite{IR2005,Glänzel}. As scientific problems grow in complexity, IDR has emerged as a promising approach for addressing societal challenges and stimulating scientific innovation \cite{Ledford2015,Rylance2015}. Studies have suggested a steady rise of interdisciplinary citations \cite{van2015,zhou2022,szell2018,Porter2009,Gates2019}, confirming the growing importance of IDR. The studies on IDR trends and patterns primarily focused on the impact of IDR on grant \cite{Bromham2016,buyalskaya2021},  researchers' interests \cite{Battiston2019,zeng2019}, and publication impact \cite{Levitt2008,Abramo2017,jing2021}. Quantifying the interdisciplinarity, diversity, variety, or balance of IDR through interdisciplinary citation is the primary method employed in literature \cite{Rafols2010,Brillouin1956,steele2000,stirling2007,lydesdorff2018, PORTER2007,Leydesdorff2011}. These studies provide insights into how IDR influences research outcomes and scientific progress. Recently, Ke et al. \cite{Qing2023} analyzed 7 million papers between 1980 and 2007 using network-standardized metrics to uncover the dynamics of high-impact works across disciplines. Understanding the developmental phases, onset, and shifting focus of IDR is crucial for advancing scientific knowledge and fostering innovation. However, large-scale, comprehensive citation networks across disciplines remain under-explored. This study aims to address the following research questions to fill this gap.

\noindent \textbf{RQ1.} What are the distinct phases in the historical evolution of interdisciplinary research (IDR), and how can we identify key developmental milestones across a large-scale dataset?

\noindent \textbf{RQ2.} What factors have contributed to the initiation and development of large-scale IDR, and which events have acted as catalysts for significant shifts? 

\noindent \textbf{RQ3.} How have the leading fields and the balance of contributions among disciplines changed over time, and what does this reveal about the focus of IDR?

To address the questions surrounding the evolution of interdisciplinary research (IDR), we introduce the concept of the Critical Year for Interdisciplinary Citations (CYIC), which identifies the timing and disciplines involved in pivotal IDR. Specifically, CYIC is defined as the year when a significant shift occurs in the citation dynamics between two disciplines, characterized by two key features: (i) the transition from one-way citations to reciprocal exchanges and (ii) a notable surge in bidirectional citations. 

\section{Methods}\label{sec2}

To systematically identify CYICs, we propose a novel metric based on two components: Interdisciplinary Balance (IB) and Knowledge Flow (KF). The IB index measures the balance of bidirectional citations between two disciplines and is defined as:

\begin{align}
IB &= 1 - \frac{|IR - IC|}{\max(IR, IC)} \label{eqn:IB_formula}
\end{align}

where IR is the number of references from another discipline, and IC is the number of citations in another discipline. This index captures the transition from one-sided influence to balanced IDR.

The Knowledge Flow (KF) quantifies the overall volume of bidirectional citations and is computed as:
\begin{align}
KF &= \frac{1}{2} IR + \frac{1}{2} IC \label{eqn:KF_formula}
\end{align}

where equal weighting is given to citations received and given. To identify CYICs, we combine these two metrics and denote the product \(z_{t,c}=IB_{t,c}\times KF_{t,c}\) for each year \(t\) across unique subject pair combinations \(c\), which serves as a measure of both the balance and intensity of interdisciplinary interactions for yearly and discipline combination values. 

CYICs are identified based on the following three conditions simultaneously:
\begin{enumerate}
\item We only consider subject combinations whose mean value is greater than the median value of all subject combinations:
\[
\frac{1}{n} \sum_{t=1}^{n} z_{t,c} > \mathrm{Median}_{t,c}(z_{t,c})
\]
where \( z_{t,c} \) represents the value of subject combination \( c \) at time \( t \), and \( n \) is the total number of years. The left-hand side represents the mean of \( z_{t,c} \) for subject combination \( c \), while the right-hand side represents the median of all subject combinations and time periods. This excludes low-value subject combinations.
\item We identify anomalous changes where the slope for the year \( \tau \) exceeds twice the standard deviation of the time series for the same subject combination:
\[
\frac{d}{dt} (z_{\tau, c}) > 2 \cdot \sigma_{z_c}
\]
where \( \frac{d}{dt} (z_{\tau, c}) \) represents the rate of change (slope) of the value \( z_{\tau, c} \) for year \( \tau \), and \( \sigma_{z_c} \) represents the standard deviation of the time series for subject combination \( c \) over the 40 years. This condition is used to detect unusually steep changes in the time series.
\item We consider the year \( \tau \) if the value of \( z_{\tau, c} \) is greater than the mean of the time series for the same subject combination:
\[
z_{\tau, c} > \mathrm{Mean}_{t} (z_{t, c})
\]
where \( z_{\tau, c} \) represents the value for year \( \tau \) and subject combination \( c \), and \( \mathrm{Mean}_{t} (z_{t, c}) \) represents the mean value of the time series for subject combination \( c \).

\end{enumerate}
By applying these criteria, CYICs indicate a marked surge when subject combinations generate significant IDR. In general, CYICs occur between subjects without prior stable citations but generate a significant amount of bidirectional citations within a short period. This metric will help to pinpoint not only when critical IDR emerges but also which subject combinations contribute most significantly to its development. This approach was validated using a dataset comprising 2,756,853 Chinese papers and 54,561,258 citations spanning 106 disciplines from 1992 to 2022\cite{Rong2024}.

\section{Results}\label{sec3}

We considered publication and citation data from Web of Science (WoS). Our dataset includes 63,092,811 papers from 254 subjects from 1981 to 2020, along with their 973,167,157 citation records. To analyze IDR trends from a broader perspective, we adapted and refined the methodology outlined in previous studies \cite{Chen2023}, clustering the 254 WoS subjects into 21 discipline clusters. Further details on the dataset, clustering method, and modifications are provided in the Supplementary information (SI).

\begin{figure*}[h]
\centering
\includegraphics[width=\textwidth, height=0.55\textwidth]{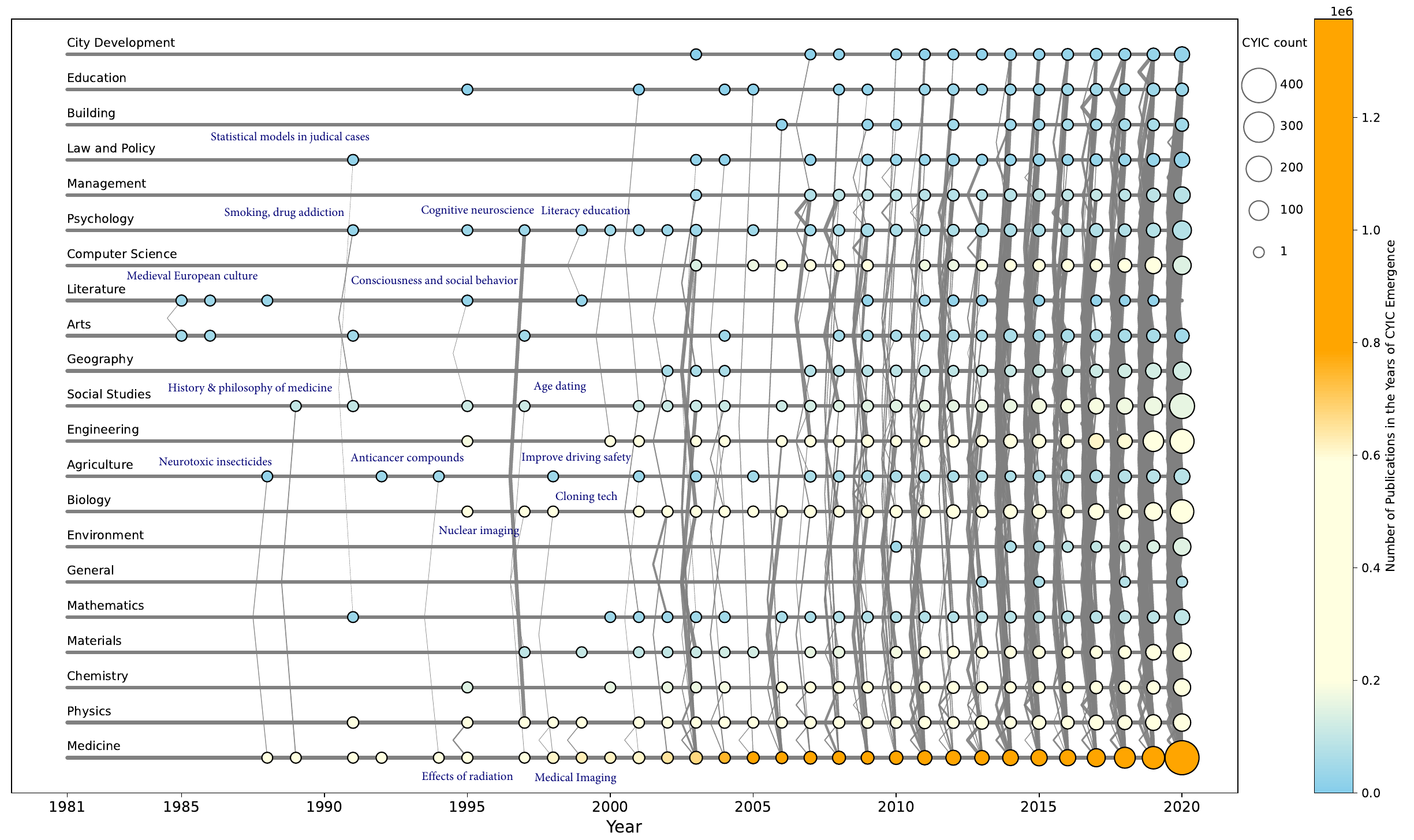}
\caption{The CYICs trends among 21 clusters from 1981 to 2020}
\label{fig_cyic_plot}
\end{figure*}

We paired the 254 WoS subjects to create 32,034 unique subject combinations with citations over the 40-year period, from which a total of 2,747 CYICs were identified. Figure \ref{fig_cyic_plot} illustrates the historical evolution of CYICs over this period, each horizontal line represents a cluster's timeline, with the cluster name labeled. Circles on the timelines indicate the occurrence of CYICs in a given year, with their size representing the number of CYICs and their color reflecting the publication counts for that year. Lines connect circles represent cross-cluster CYICs and are annotated with events or technological advancements that triggered early IDR. The thickness of these lines corresponds to the average \( z_{t,c} \) value of the cross-cluster CYICs. Circles without connections indicate CYICs that occurred solely within the same cluster that year. Among these, 2,529 CYICs were cross-cluster, represented by lines linking two discipline clusters, accounting for 92.06\%, while 218 CYICs occurred between subjects within the same discipline cluster, highlighted as circles without lines.

\subsection{The development phases of IDR}\label{subsec1}

We segmented the IDR into three periods based on the annual number and cluster number of CYICs. In Figure \ref{fig_cyic_plot}, 2003 marks the first turning point, as there was a significant surge in both annual cross-cluster CYICs and the participating clusters. Specifically: (1) the annual cross-cluster CYICs reached 24, compared to an average of 2.5 and a maximum of 10 in the prior 22 years; (2) the annual number of clusters involved in cross-cluster CYICs rose to 13, whereas the annual average before 2003 was 2, with a maximum of 6. These statistics indicate that the scale of IDR experienced significant growth in 2003, both in terms of the number of CYICs and the participating clusters. The second significant turning point occurred in 2017, marking the first instance of a growth rate exceeding 50\% when the base number of annual cross-cluster CYICs was over 100. In particular, the number of cross-cluster CYICs rose from 125 in 2016 to 206 in 2017, reflecting a 64.8\% increase. Before this, the average growth rate in years where the base number exceeded 100 was only 2.8\%. From 2017 to 2020, however, the average growth rate reached 64.68\%, with a maximum of 109.59\%. This data indicates that after 2017, the number of cross-cluster CYICs experienced a substantial surge.

\begin{table}[h]
\centering
\setlength{\tabcolsep}{3pt}  
\small 
\caption{Clusters by Cross-Cluster CYICs in Each Period}
\label{tab_Sub_3period}
\begin{tabular}{c l l l}
\toprule
Rank & \multicolumn{1}{c}{\shortstack{1981--2002 \\ Cluster (Count)}} & \multicolumn{1}{c}{\shortstack{2003--2016 \\ Cluster (Count)}} & \multicolumn{1}{c}{\shortstack{2017--2020 \\ Cluster (Count)}} \\
\midrule
1  & Medicine (30)         & Medicine (331)         & Medicine (616)         \\
2  & Physics (16)          & Biology (168)          & Engineering (345)      \\
3  & Biology (11)          & Social Studies (141)   & Social Studies (313)   \\
4  & Social Studies (9)    & Physics (138)          & Biology (309)          \\
5  & Psychology (8)        & Engineering (112)      & Physics (186)          \\
6  & Agriculture (7)       & Psychology (110)       & Computer Sci. (182)    \\
7  & Materials (5)         & Materials (71)         & Geography (170)        \\
8  & Mathematics (4)       & Agriculture (71)       & Psychology (158)       \\
9  & Engineering (4)       & Management (66)        & Materials (157)        \\
10 & Literature (4)        & Computer Sci. (65)  & Chemistry (124)           \\
11 & Arts (4)              & Arts (59)              & Agriculture (116)      \\
12 & Chemistry (4)         & Geography (58)         & Management (101)       \\
13 & Geography (2)         & Chemistry (53)         & Environment (100)      \\
14 & Education (2)         & Mathematics (52)       & Arts (87)              \\
15 & Law and Policy (1)    & Law and Policy (42)    & Mathematics (86)       \\
16 & City Development (0)  & City Development (30)  & City Development (82)  \\
17 & Management (0)        & Education (25)         & Law and Policy (82)    \\
18 & Computer Sci. (0)     & Building (18)          & Education (54)         \\
19 & Environment (0)       & Environment (12)       & Building (44)          \\
20 & General (0)           & Literature (4)         & General (4)            \\
21 & Building (0)          & General (2)            & Literature (4)         \\
Total  & Unique CYICs (55) & Unique CYICs (814)     & Unique CYICs (1660)    \\
Avg.   & Per year (2.5)    & Per year (58.14)       & Per year (415)         \\
\bottomrule
\end{tabular}
\end{table}

Therefore, we used 2003 and 2017 as division points to segment the development of IDR into three period: Period I (1985–2002), Period II (2003–2016), and Period III (2017–2020). Table \ref{tab_Sub_3period} presents the rankings of cross-cluster CYIC counts for each cluster across the three periods. The number of cross-cluster CYICs varied widely across the three periods, averaging 2.5 per year in Period I, rising to 58.14 in Period II, and surging to 415 in Period III.

\textbf{Medicine stands out as the top performer across all three periods, highlighting its leader role in IDR.}  Physics, Biology, and Social Studies have consistently ranked among the top clusters, underscoring their substantial contributions to IDR. Although the number of cross-cluster CYICs in Psychology, Agriculture, Materials, Mathematics, and Education continued to grow, their rankings declined. The rankings of Engineering, Environment, and Geography improved, with Engineering making notable progress, advancing from ninth place in Period I to second place in Period III. These changes indicate that the characteristics of IDR were not static across the three periods.

\subsubsection{Period I (1981--2002)}\label{subsubsec1}

IDR was sporadic, limited to a few clusters, and had not yet evolved into large-scale. IDR were often driven by social issues or technological advancements in specific fields in this period. In Table \ref{tab_Sub_3period}, the number of cross-cluster CYICs and participating clusters is the lowest among the three periods, with six clusters showing none over the 22 years.

In Figure \ref{fig_cyic_plot}, the connections between clusters are very sparse, with an average of only two clusters per year exhibiting cross-clusters CYICs. \textbf{The earliest IDR emerged in Literature and Arts in 1985.} However, their cross-cluster IDR were less frequent compared to their internal cluster citations. Medicine and Agriculture were identified with CYIC twice: in 1988, driven by the development of neurotoxic insecticides in response to pest outbreaks, and in 1994, with a primary focus on anticancer compounds extracted from plants. Social Studies engaged in the most diverse IDR with other clusters, sequentially forming cross-cluster CYICs with medicine, psychology, literature, education, and mathematics. These IDR were triggered by various research advancements or events, such as studies on the history of medicine, societal issues like smoking, drug addiction, and carcinogenesis, as well as topics like consciousness studies, and isotopic age measurement on archaeological research. In 1991, Mathematics and Law were identified with CYICs, as statistical models were applied to judicial cases to examine the influence of various factors on trial outcomes. Psychology and Physics recorded the highest mean \( z_{t,c} \) in 1997 during this period, represented in Figure \ref{fig_cyic_plot} by the thickest cross-cluster connection, primarily collaborating on research in cognitive neuroscience.

\subsubsection{Period II (2003--2016)}\label{subsubsec2}

This phase differed from Period I in two aspects: (1) all clusters gradually began to participate in IDR; (2) most clusters' IDR began to show continuity. 

In Figure \ref{fig_cyic_plot}, the number of CYICs increased steadily each year, and the connections between clusters became progressively denser, with most clusters exhibiting continuous circles. Only 4 clusters were exceptions to this trend. Literature was the only cluster to show a declining trend in both publication and CYIC. The discontinuous trend observed in Computer Science may be attributed to a preference for presenting research progress at conferences, whereas our dataset focuses on journal articles. Environment had very few papers during this period. As for General, most publications appeared in comprehensive journals that consistently engaged in extensive and stable IDR. Consequently, the frequency and balance of their interdisciplinary citations remained relatively stable, resulting in fewer identified CYICs. Although these 4 clusters did not display a continuous growth trend before 2016, they still demonstrated CYIC during this period, suggesting that IDR had emerged across all clusters.

\subsubsection{Period III (2017-- )}\label{subsubsec3}

Since 2017, the number of cross-cluster CYICs has experienced exponential growth, with over 65\% occurring during this period. Compared to the previous two periods, the main changes were: (1) IDR had become a common research mode; (2) the top clusters of IDR exhibited significant changes.

In Figure \ref{fig_cyic_plot}, IDR had reached an extensive scale, with IDR between clusters becoming increasingly complex during this period. All clusters show a clear trend toward large-scale IDR, except for Literature and General. During this period, the annual number of cross-cluster CYICs increased by 613.94\%, while the annual total number of citations increased by 87.88\% compared to the previous Period II. The growth of IDR far exceeded that of total citations, indicating that IDR had become a common research mode and played a crucial role in scientific research.

Three pieces of evidence indicate the shift in the direction of IDR during this period. (1) Engineering and Social Studies became the clusters with the most CYICs with Medicine, while Biology and Physics dropped to third and fourth place. This shift indicates a change in the IDR partners of Medicine, which has remained the dominant cluster in IDR. (2) In Table \ref{tab_Sub_3period}, Engineering experienced the most significant interdisciplinary progress among all clusters across the three periods. It grew from 4 CYICs in the Period I to 345 CYICs in the Period III, ranking second among all clusters. This indicates that engineering applications are gaining increasing attention across clusters and have developed into a highly significant focus area. (3) In Figure \ref{fig_cyic_plot}, the development of the Environment is noteworthy. Although its overall performance is not outstanding, and its involvement in IDR began relatively late, it took only 7 years for its CYIC count to rise from well below the average to a level comparable with other clusters.

\subsection{The beginning of large-scale IDR}\label{subsec2}

We found that large-scale IDR began in 2003, as the number of cross-cluster CYICs grew steadily in Period II. The disciplines that first exhibited stable IDR were mostly related to Medicine. Breakthroughs in cloning technology served as the prelude to large-scale IDR, while advancements in medical technology were the primary driver of its emergence. 

Two pieces of evidence support that Medicine may trigger large-scale IDR. First, in Figure \ref{fig_cyic_plot}, between 1995 and 2005, cross-cluster CYICs between medicine and biology, as well as between medicine and physics, occurred in six of the ten years. Second, In 2003, IDR with Medicine was the most prominent, among the 24 cross-cluster CYICs that emerged this year, 13 were involved Medicine. These data suggest that the beginning of large-scale IDR was likely closely associated with Medicine.

Our further investigation revealed that this period coincided with the rise of genetic research, which may explain the frequent IDR between Medicine and Biology. One of the most notable technological breakthroughs of this era was the birth of Dolly in 1996\cite{Wilmut1997}, which provided both theoretical and practical proof of the feasibility of animal cloning using somatic cells. Our raw data supports this, showing that much of the interdisciplinary citations between Medicine and Biology during these ten years focused on topics such as gene cloning, knockout, modification, recombination, and transformation. The rise of genetic research also contributed to the development of physical experimental instruments, such as laser micro-manipulation technology, high-throughput sequencing, and spectral imaging. This suggests that the rise of genetic research in the late 20th century fostered IDR. Although it had not yet reached a large scale, it can be considered a prelude to modern IDR.

The research into medical technologies was the main driver that propelled modern IDR to expand on a larger scale. For example, between 1995 and 2005, the frequent IDR between physics and medicine can be attributed to studies on medical imaging, sensing, and physiological signal monitoring techniques. In 2003, during the first surge in the number of CYICs, Neurosciences and Cardiac \& Cardiovascular Systems were the most frequently occurring subjects, which primarily collaborated with fields such as physics, psychology, and biology. The IDR in the field of Neurosciences primarily focuses on neural network modeling, brain function and behavior studies, as well as interactions with external environments. The IDR in the field of cardiac primarily focuses on image modeling, cardiac-related technologies, as well as cardiac dynamics and complex system analysis. 

Medicine also played a pivotal role in the development of IDR. Across the three periods, 1,077 cross-cluster CYICs were related to medicine, accounting for 42.59\% of the total. As evidenced in Table \ref{tab_Sub_3period} and Figure \ref{fig_cyic_plot}, medicine consistently ranked first in all periods, with the number of CYICs significantly surpassing those of the second-ranked cluster.

\begin{figure}
\centering
\includegraphics[width=\linewidth]{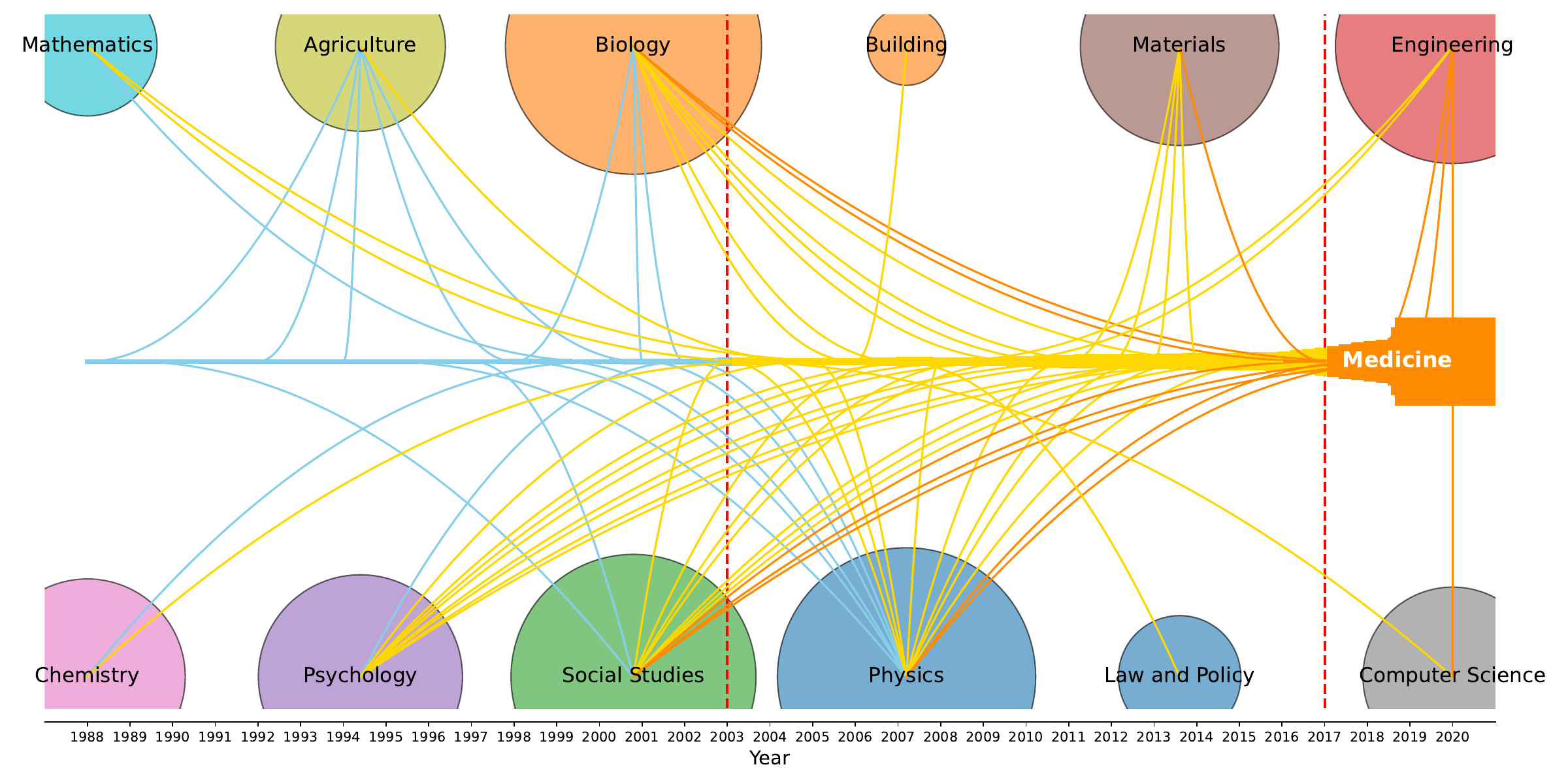}
\caption{Evolution of Medicine’s Top 3 IDR Partnerships}
\label{fig_Sg_medicine} 
\end{figure}

Figure \ref{fig_Sg_medicine} illustrates the changes in IDR clusters with Medicine across three periods. The central main line represents "Medicine," with its thickness indicating the frequency of CYICs. Circles representing clusters that ranked in the Top 3 for annual CYICs with Medicine. The size of each circle reflects the total number of CYICs generated with Medicine. The lines connecting the circles to Medicine indicate the years of CYICs. The main line and connecting lines are highlighted in different colors, the blue line shows IDR from Period I, the yellow from Period II, and the orange from Period III.

(1) Medicine maintained continuous and stable IDR with the Physics, Biology, and Social Studies. This is evident from the relatively even distribution of connections with the Medicine. Physics stands out as a consistently stable interdisciplinary collaborator with Medicine, ranking in the top three for cross-cluster CYICs with Medicine in 15 out of the 40 years. (2) Mathematics, Chemistry, Agriculture, and Psychology engaged in CYICs with Medicine exclusively during Periods I and II. Notably, Agriculture was more active in Period I, while Psychology showed greater involvement in Period II. Building and Law and Policy entered the top three only once during Period II. These patterns suggest that as Medicine's IDR has developed, its IDR partners have gradually shifted. (3) Materials, Engineering, and Computer Science appeared in the top three collaborators with Medicine only during Periods II and III. Among these, Engineering had the most occurrences in Period III, Materials collaborated more frequently with Medicine in Period II, and Computer Science entered the top three once in each of the two periods.

These data suggest that although Medicine's core IDR with Physics, Social Studies, and Biology have remained stable, its other IDR focuses are gradually shifting from fields such as Agriculture, Psychology, and Mathematics to cluster like Engineering and Materials. Moreover, following the onset of large-scale IDR, citations between Agriculture and Medicine has gradually faded from prominence.

\subsection{Changes in the focus of IDR}\label{subsec3}

The data suggest that, compared to discipline clusters in Humanities \& Social Sciences (e.g. Law and Policy, Arts), IDR is more active within discipline clusters in Natural Sciences (e.g. Engineering, Medicine). The majority of IDR growth has been observed within the Natural Sciences, whereas in Humanities \& Social Sciences, only Social Studies has shown significant growth. The decline in IDR activity is predominantly observed at the intersection of Humanities and Social Sciences with Natural Sciences. Furthermore, while Medicine remains the dominant cluster in IDR, the focus is gradually shifting towards Engineering and Environment, as citations with them are increasing across all clusters.

\begin{figure}
\centering
\includegraphics[width=\linewidth]{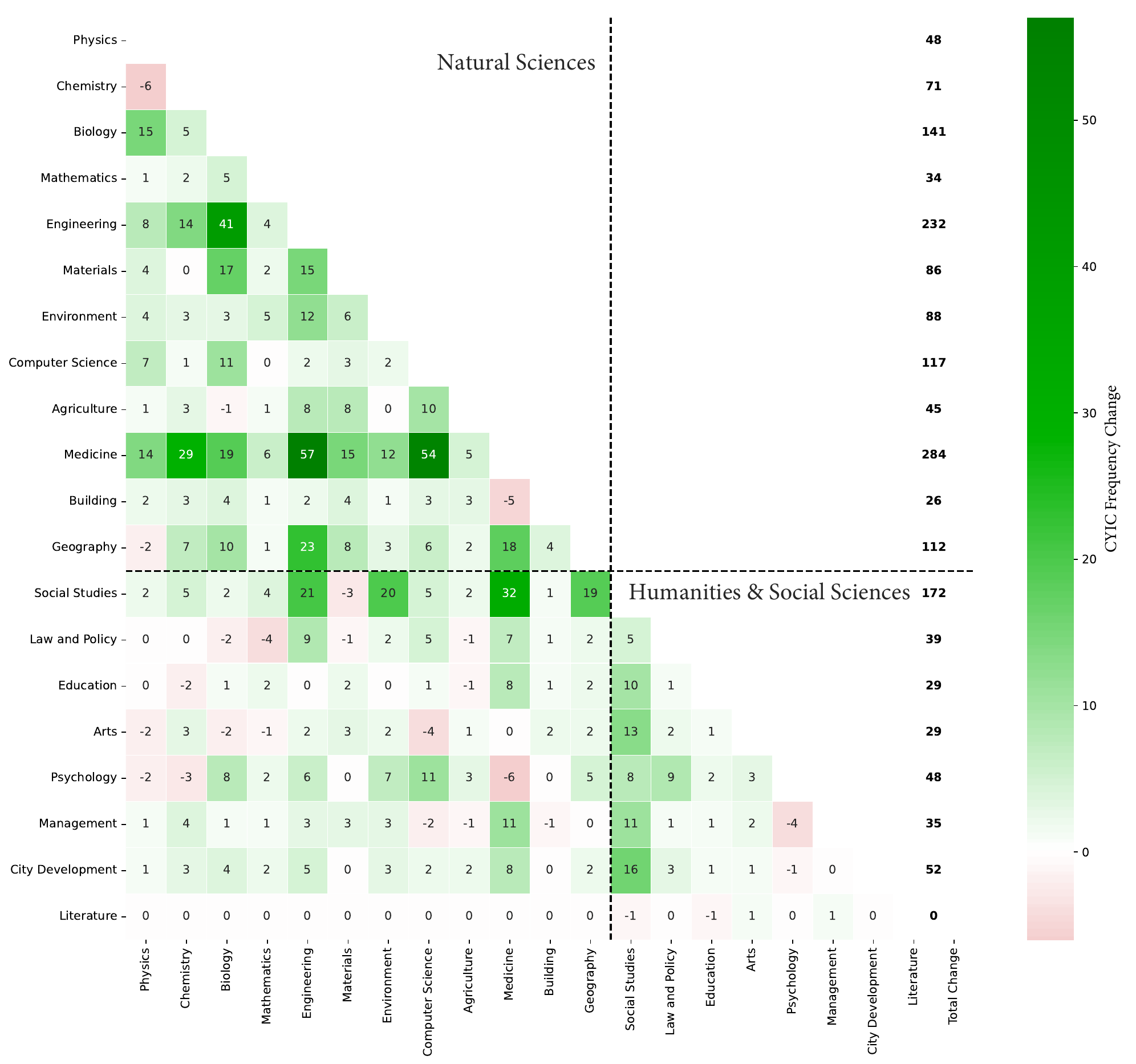}
\caption{Changes in cross-cluster CYICs: Period III vs. Period II}
\label{fig_matrix}
\end{figure}

Figure \ref{fig_matrix} illustrates the changes in CYICs for each cluster in Period III compared to Period II (without General). The values in the matrix represent the differences between the cross-cluster CYICs of the two periods. The total change for each cluster is marked in the last column of the matrix. To facilitate comparison, we distinguish between Natural and Humanities \& Social Sciences. The upper-left quadrant shows changes in CYICs within the Natural Sciences, the lower-right within the Humanities \& Social Sciences, and the lower-left between the two. Overall, the number of CYICs exhibits an upward trend, as indicated by the fact that all cluster's total change are greater than or equal to zero. In the three sections of Figure \ref{fig_matrix}, the number of CYICs within the Natural Sciences exhibits the highest growth, with an average increase of 7.95 CYICs per cluster combination, compared to the overall average of 4.44. Furthermore, 89.39\% of cluster combinations within the Natural Sciences show an increase in CYICs, compared to an overall rate of 71.05\%. This indicates that IDR is most active within the Natural Sciences. In contrast, the intersection of Natural Sciences and Humanities \& Social Sciences exhibits the lowest level of IDR activity, with an average increase of only 2.44 CYICs per cluster combination, and 46\% of cluster combinations showing a decline in CYICs. These data suggest that the focus of IDR is increasingly centered on clusters within the Natural Sciences.

The vibrancy of IDR within the Natural Sciences is primarily driven by Medicine and Engineering, while the rare instances of declining are mainly associated with Physics, Chemistry, Building, and Geography. Medicine was the cluster with the highest growth in IDR, contributing the largest number of new CYICs with Engineering and Computer Science. This reinforces the dominant role of medicine in IDR. However, IDR between Medicine and cluster such as Building and Psychology have noticeably weakened, as evidenced by a decline in the number of CYICs with Medicine. This downward trend is particularly striking within the context of the overall growth, indicating that the IDR activity of a few cluster with medicine is declining. Additionally, the number of CYICs between Physics and clusters such as Chemistry and Geography has also declined. Combined with the continuous drop in Physics' ranking shown in Table \ref{tab_Sub_3period}, this suggests that the IDR activity of Physics is diminishing. In contrast, Engineering and Environment have shown more comprehensive progress in IDR, with no decline observed across all clusters. Engineering, ranking second only to Medicine in growth, has achieved significant progress in IDR through citations with Medicine, Biology, and Geography. 

In Humanities \& Social Sciences, Social Studies remains the primary contributor to IDR, while other clusters have shown little increase in participation. Notably, citations between Psychology, Education, Arts, and Law and Policy with certain clusters of the Natural Sciences have experienced significant declines, Literature has consistently exhibited minimal IDR activity. Social Studies ranks third among all clusters in terms of CYIC growth, with its primary IDR focused on Natural Sciences such as Medicine, Engineering, and Environment. Its CYIC growth with other Humanities \& Social Sciences clusters, such as City Development and Arts, is lower than that with the Natural Sciences. This data indicates that Social Studies tends to collaborate more actively with the Natural Sciences, with Materials being an exception. 

However, the IDR trends between other clusters of Humanities \& Social Sciences with Natural Sciences appear less encouraging. Arts shows the most widespread decline in CYICs with Natural Sciences, experiencing varying degrees of decrease with Physics, Biology, Mathematics, and Computer Science. Psychology has the highest number of CYIC declines, primarily in citations with Medicine, Physics, and Chemistry.  Education, Management, and Law and Policy have also experienced slight declines. Literature, having low citations with the Natural Sciences, shows no observable changes. Furthermore, changes in IDR within the Humanities \& Social Sciences clusters are minimal. Apart from IDR between Psychology and Law and Policy, no significant changes have been observed.

\section{Discussion}\label{sec4}
Our study addressed three research questions and led to three conclusions. (1) IDR has undergone three distinct developmental phases. Period I (1981--2002) was sporadic, limited to a few clusters, and had not yet evolved into large-scale. During Period II (2003–2016), all clusters gradually began to engage in IDR, with most clusters demonstrating increasing continuity in their IDR activities. By Period III (2017–present), IDR had become a common research paradigm, accompanied by significant changes in the leading clusters of IDR. (2) Large-scale IDR began in 2003, led by Medicine, with cloning technology serving as a prelude, while advancements in medical technology were the primary driver. Medicine has consistently led IDR and has undergone shifts in its IDR partners as it has evolved. (3) IDR is more active within the Natural Sciences, where Medicine maintains its dominant position, alongside a growing focus on Engineering and Environment.

\subsection{Possible mechanisms behind the rise of IDR}\label{subsec4}

We can infer from the results that early research mainly focused on discipline-specific approaches, with fragmented IDR sparked by major social events and technological advances. During this time, IDR was more problem-oriented, resulting in numerous classic achievements. After more than two decades of technological progress, significant breakthroughs that drew widespread attention emerged at the end of the 20th century. These breakthroughs acted as a starting point for large-scale IDR, which began to grow under the influence of Medicine in the early 21st century. It implies that IDR requires significant prior groundwork. Large-scale IDR typically begins when theories and technologies within individual fields become sufficiently developed. This enables researchers to tackle problems that single disciplines cannot solve or to explore new paths for combining technologies. It is not surprising that medicine, with its extensive prior knowledge and practical experience, was the first to initiate IDR.

With the rise of large-scale IDR, the motivations behind IDR may have shifted. In Figure \ref{fig_cyic_plot}, the most notable evidence is that, although large-scale IDR began to emerge in the early 21st century, pinpointing a clear trigger event or specific technology that initiated it has become challenging. Moreover, the main interdisciplinary problems IDR seeks to address have become more ambiguous. In Period I, IDR addressed numerous practical problems, resulting in benefits that continue to impact the world today. Notable examples include neurotoxin-based insecticides, anticancer compounds, the causes of smoking and drug addiction, radiation effects, medical imaging technology. These achievements demonstrate that IDR is an important means of addressing societal and scientific problems, and they have also encouraged more scientists to engage in IDR. As the scale of IDR expanded and the boundaries between disciplines became increasingly blurred, it has become difficult to identify the main themes and research questions of IDR. This shift reflects the transition of IDR from a problem-oriented approach focused on societal issues and technological development to a research mode centered on methodological transfer.

Advancements in technology and data accessibility may have also driven the development of large-scale IDR, as they have facilitated the transfer of methods across disciplines. New tools, shared data platforms, and computational models have made it easier for scientists to access cross-disciplinary data and apply methods from different fields. This not only lowered the cost of study for conducting IDR but also provided new data, technologies, and solutions for longstanding research questions. The lower barriers spurred an increase in research utilizing IDR for methodological and data transfer. While such studies address Scientific problems from new perspectives, they inevitably reduced the proportion of problem-oriented IDR. This may also explain why changes in IDR are more notable in the Natural Sciences. The Natural Sciences often serve as the source of new technologies and methods, while the Humanities \& Social Sciences typically focus on addressing societal issues. Data and technological exchanges are more frequent within the Natural Sciences, leading to the highest increase in CYICs.

Current trends indicate that IDR will continue its explosive growth as a widespread research mode. We speculate that Engineering will continue to show a rising development trend after 2020, and Environment having the greatest potential for growth. Medicine is likely to further widen the gap with other clusters. Our dataset ends in 2020, the impact of the COVID-19 had not yet been fully reflected in academic publications and citations at that time. It is unfortunate that we are unable to assess the influence of this significant global event on IDR.

\subsection{Limitations and future research}\label{subsec5}

Although CYIC can detect anomalous fluctuations in interdisciplinary citations, it cannot systematically quantify the temporal delay between citation surges and major real-world events, due to the inherent time lag in publication and citation processes. Consequently, our interpretation of CYIC drivers primarily relies on the research topics identified in the raw data, supplemented by supporting evidence from relevant literature. It should be emphasized that while the CYIC enables accurate detection at the citation level, revealing its underlying mechanisms still necessitates modest inferential analysis. Future studies will strengthen the linkage between CYIC and historical events to better assess the impact of major occurrences on IDR.

Our findings are constrained by dataset limitations. Since Web of Science (WoS) rarely indexes conference proceedings and non-English publications, the current analysis only reflects interdisciplinary trends in English journal articles. Future research will expand the dataset to provide a more comprehensive understanding of IDR patterns at a macro level.

While our method captures the overall trends in IDR, it currently does not incorporate citation normalization across disciplines of varying sizes. Current approach offers the advantage of providing a macro-level perspective on IDR patterns, particularly in identifying fields with both high IDR volume and significant citation fluctuations. However, it may inadvertently diminish the visibility of smaller disciplines that demonstrate proportionally higher IDR engagement despite their lower publication output. To address this limitation, future work will develop a discipline-size-adjusted normalization method for cross-disciplinary citations and integrate it into the CYIC framework to enhance analytical robustness.

\backmatter

\bmhead{Supplementary information}

Methodological details, including dataset description, data processing procedures, clustering methodology, identification cases, and case analysis approaches, are comprehensively documented in Supplementary Information (SI). Additionally, we have made process data, result data, and computational code publicly available, with access details provided in SI.

\bmhead{Acknowledgements}

Part of the work has been co-funded by the European Union (European Regional Development Fund ERDF, fund number: STIIV-001 ). Part of the work has been conducted in the Research Campus MODAL funded by the German Federal Ministry of Education and Research (BMBF) (fund numbers 05M14ZAM, 05M20ZBM). The authors thank Qingqing Fan, Yujiao Sun, and Jinglong Chen for their assistance in collecting the literature on the events labeled in Figure \ref{fig_cyic_plot}.

\bmhead{Author contribution}

Guoyang Rong and Ying Chen designed the research; Guoyang Rong performed the research; Guoyang Rong analyzed the data; Guoyang Rong and Ying Chen wrote the paper; Ying Chen, Feicheng Ma, and Thorsten Koch provided critical revisions.

\bmhead{Competing interests}

The authors declare no competing interest.


\bibliography{sn-bibliography}

\end{document}